\documentstyle[preprint,aps,psfig]{revtex}

\tighten
\begin{document}
\draft
\preprint{Imperial/TP/96-97/69, DAMTP-97-106}

\newcommand{\nc}{\newcommand}
\nc{\al}{\alpha}
\nc{\ga}{\gamma}
\nc{\de}{\delta}
\nc{\ep}{\epsilon}
\nc{\ze}{\zeta}
\nc{\et}{\eta}
\renewcommand{\th}{\theta}
\nc{\ka}{\kappa}
\nc{\la}{\lambda}
\nc{\rh}{\rho}
\nc{\si}{\sigma}
\nc{\ta}{\tau}
\nc{\up}{\upsilon}
\nc{\ph}{\phi}
\nc{\ch}{\chi}
\nc{\ps}{\psi}
\nc{\om}{\omega}
\nc{\Ga}{\Gamma}
\nc{\De}{\Delta}
\nc{\La}{\Lambda}
\nc{\Si}{\Sigma}
\nc{\Up}{\Upsilon}
\nc{\Ph}{\Phi}
\nc{\Ps}{\Psi}
\nc{\Om}{\Omega}
\nc{\ptl}{\partial}
\nc{\del}{\nabla}
\nc{\be}{\begin{eqnarray}}
\nc{\ee}{\end{eqnarray}}
\nc{\lambar}{\overline{\lambda}}
\nc{\cm}{{\cal M}}
\nc{\cah}{{\cal H}}


\title{On the Symmetry of Real-Space Renormalisation} 
 
\author{
Dorje C. Brody\footnote[1]{Electronic address: d.brody@damtp.cam.ac.uk}
and 
Adam Ritz\footnote[2]{Electronic address: a.ritz@ic.ac.uk} 
} 
\address{$*$Department of Applied Mathematics and 
Theoretical Physics,  \\ University of Cambridge,  
Silver Street, Cambridge CB3 9EW U.K.} 
\address{$\dagger$ Blackett Laboratory, Imperial College, 
South Kensington, London SW7 2BZ U.K.}  

\date{\today}
\maketitle

\begin{abstract} 
A natural geometry, arising from the embedding into a Hilbert space 
of the parametrised probability measure for a given lattice model, 
is used to study the symmetry properties of real-space 
renormalisation group (RG) flow. In the projective state space this 
flow is shown to have two contributions: a gradient term, which generates
a projective automorphism of the state space for each given length scale;
and an explicit correction. We then argue 
that this structure implies the absence of any symmetry of a geodesic type
for the RG flow when restricted to the parameter space 
submanifold of the state space. This is demonstrated explicitly via a study 
of the one dimensional Ising model in an external field. In this 
example we construct exact expressions for the beta functions associated 
with the flow induced by infinitesimal rescaling. These constitute 
a generating vector field for RG diffeomorphisms on the 
parameter space manifold, and we analyse the symmetry properties of 
this transformation. The results indicate
an approximate conformal Killing symmetry near 
the critical point, but no generic symmetry of the 
RG flow globally on the parameter space.  
\end{abstract} 



\section{Introduction}

The construction of effective theories for the low energy modes
from their underlying microscopic dynamics
or, equivalently, the study of the thermodynamic behaviour of spin and 
lattice field systems 
is a fundamental problem in modern physics. The standard theoretical
framework for this procedure, the renormalisation group 
\cite{wil}, provides a methodology for systematically integrating 
out high momentum or short distance modes leaving an effective 
dynamical system for the long wavelength physics. This effective theory
then determines, in particular, the critical behaviour of field theories or 
alternatively the thermodynamic behaviour of lattice 
or statistical mechanical spin systems. \par

The fundamental obstacle in following this procedure through to 
completion is that coarse graining the system, regardless of the 
detailed approach employed, generically leads to a highly complex 
effective theory introducing an often arbitrarily high number of 
additional interactions. While some simplifications occur near the 
upper critical dimension for a given system, allowing systematic 
techniques such as the $\ep$--expansion to be employed, generically 
one requires drastic approximations in order to render the RG 
procedure practically implementable. In particular, we focus on real-space 
renormalisation \cite{bl} methods which implement an intuitive blocking 
picture in attempting to explain certain universal aspects of 
critical behaviour, often near the lower critical dimension. 
However, except in a few tractable cases, such calculational schemes 
involve significant truncations, and there is rarely a useful 
criterion for selecting a particular approximation. This is 
also a significant problem in more recent attempts to study field 
theories via Wilsonian RG techniques \cite{rgtech} as the dominant infrared 
degrees of freedom in such cases may differ greatly from the relevant  
microscopic fields. \par 

With these problems in mind it is clearly of interest to study the 
structure of the RG in some detail. Generically the Hamiltonian for 
a spin system, or the Wilsonian action for a field theory, at a 
given scale may be represented in the form, $H=\sum_i \th^i H_i$. This is
generally an infinite series of interaction Hamiltonians $H_i$ and 
their associated scale dependent couplings $\th^i(t)$. 
The RG acts on these couplings as one coarse grains the system, and 
one may gain insight into this process by studying the geometry of 
the parameter space, whose singularity structure and possible 
symmetries determine the properties of the RG flow, in particular 
the fixed points. However, the manifold $\cm$ of the parameter set 
dealt with here is typically a subspace of an infinite dimensional 
Hilbert space $\cah$ \cite{bh} and this infinite dimensionality, 
which in effect is the cause of the singularities, gives rise to 
various technical problems that have to be treated with caution. 
Nevertheless, the ability to formulate certain RG related questions 
in a geometric formalism allows the utilisation of various powerful 
techniques in differential geometry. \par 

The Riemannian structure of this parameter manifold $\cm$, as we 
discuss in more detail below, is inherited from the structure 
of the probability measure of the system over its configuration space
\cite{bh}. Various authors \cite{diosi,geo,janyszek89,stat} have 
utilised this structure to study aspects of statistical mechanical 
systems and field theories, and it has been shown that \cite{geo} 
the RG acts via diffeomorphisms of this parameter space
manifold generated by the $\beta$--function vector field. It is then 
a clear corollary that the geometry can place restrictions on the 
$\beta$--function of the theory if other symmetries of the 
parameter space exist, or indeed if the RG flow itself is a symmetry 
of ${\cal M}$.\par 

In this paper, we investigate this last question, and consider 
whether the RG diffeomorphism actually corresponds to a symmetry of 
the parameter space. This is motivated by earlier work 
\cite{diosi,geodesic} which suggests that an approximate symmetry 
holds near the critical points, and also to some extent by recent 
work \cite{dblh2,dblh3} on the geometric structure of the state space 
of statistical mechanical systems, and in particular the parametric 
evolution of the states. A consequence of this latter work is that 
for a system whose action is expressible in the form $\sum_i \th^i H_i$,
the parametric evolution of the state corresponds to a projective automorphism
of the projective Hilbert space, which is the most general 
transformation of a Riemannian manifold mapping geodesics into 
geodesics.\par 

In generalising this result to systems where
all the parameters depend on a single dimensionful
scale, we show in Section 2 that the generating vector field of
the flow, when restricted to a {\it constant} scale, still generates
a projective automorphism of the state--space. However, in considering the
full scale dependence of the states, it is shown that
the integral curve corresponding to RG evolution
only generates a projective automorphism of the state-space
up to an anomalous correction. Since the parameter space 
can be viewed as a submanifold of the projective state-space,
this suggests the absence of such a geodesic type symmetry 
of the RG flow on the parameter space itself. The generality of this 
construction allows us to conjecture that
in general the RG does {\it not} generate any standard
global symmetry of this type on the parameter space. \par 

In order to test this conjecture we provide an explicit counter-example
in which both the RG trajectory, and the underlying parameter space
geometry, can be exactly determined. While the RG procedure 
generally requires various approximations, there are a number of 
exactly tractable models known in statistical mechanics 
\cite{janyszek89,stat}, and in the present paper we study one such 
model exhaustively by use of real-space RG methods, in order to 
determine the properties of RG flow in the relevant parameter 
space. \par 

The example we consider here is the one dimensional Ising model in 
an external magnetic field. As we shall discuss in more detail below, 
the parameter space in this case is a two dimensional manifold 
endowed with a Riemannian structure in terms of the Fisher-Rao 
metric. In Section 3 we consider the exact parameter space geometry 
of this system both for a finite $N$-spin chain and in the 
thermodynamic limit. In Section 4 we utilise a transfer matrix 
method to obtain an exact expression for the vector 
field associated with an infinitesimal change of the lattice 
spacing, the components being the differential RG $\beta$-functions.
Using this result we study, in Section 5, the symmetry associated 
with the flow. We find, in particular, that the generating vector 
field corresponds to a conformal Killing vector field in the 
vicinity of the critical point. However, this vector field does 
not generate such a transformation, nor the most general additional 
class of mappings, a projective automorphism, globally on 
the manifold. Therefore we are led to conclude, via this explicit
counter-example, that there is no generic global symmetry structure 
of a standard form for the RG flow. We finish in Section 6 with some
concluding remarks. \par 

\section{State Space Renormalisation} 

We begin by considering the effect of renormalisation group flow in 
the actual state space of a given system, which can be viewed as the 
space of rays through the origin of a real Hilbert space ${\cal H}$, 
that is, the real projective $n$-space $RP^{n}$ (possibly infinite 
dimensional). Before studying the behaviour of RG flow in this 
space we review briefly the notion of the state space in statistical 
mechanics \cite{dblh3}. \par 

In a purely statistical mechanical context we are given the 
parametrised family of probability distributions, taking the form 
of the Gibbs measure, 
\begin{equation} 
p(x,\theta)\ =\ q(x) \exp\left[- \sum_{j}\theta^{j}H_{j}(x) 
- W_{\theta} \right] \ , \label{eq:gib} 
\end{equation} 
where the variable $x$ ranges over the configuration space, 
$H_{j}(x)$ represents the form of the energy, $W_{\theta}$ is a 
normalisation factor, and $q(x)$ determines the distribution at 
$\theta^{j}=0$. In a field theoretic context the operators $H_{j}$ 
effectively determine the form of the action, and the parameters 
$\theta^{j}$ are viewed as the coupling constants. Now, by taking 
the square-root of the distribution function (\ref{eq:gib}) we can 
map, for each given value of $\theta^{j}$, the space of probabilities 
onto a manifold in a real Hilbert space ${\cal H}$. Since the 
probability distribution is normalised, we find that for each fixed 
value of $\theta^{j}$, the probability state corresponds to a point 
on the unit sphere ${\cal S}\subset {\cal H}$. Then, by choosing a 
suitable basis of vectors in ${\cal H}$, we can consider the vector 
$\psi^{a}(\theta)$ as representing the point characterised by 
$\sqrt{p(x,\theta)}$. In other words, we regard $\psi^{a}(\theta)$ 
as a state vector corresponding to the distribution (\ref{eq:gib}). \par 

First we would like to formulate a Hilbert space characterisation of 
this distribution. In fact, it can be shown \cite{dblh2,dblh3} that 
the state vector $\psi^{a}(\theta)$ in ${\cal H}$ corresponding to 
the Gibbs distribution (\ref{eq:gib}) satisfies the differential 
equation,
\begin{equation} 
\frac{\partial\psi^{a}}{\partial\theta^{j}}\ =\ - 
\frac{1}{2}{\tilde H}^{a}_{j b} \psi^{b}\ , \label{eq:thermo} 
\end{equation} 
where the operators $H_{j ab}$ in ${\cal H}$ represent the energy 
$H_{j}(x)$, ${\tilde H}_{j ab} = H_{j ab}-g_{ab}E_{\psi}[H_{j}]$, 
with $E[\cdot]$ denoting the expectation. The solution to this 
equation is given by an exponential family \cite{bh} of states 
\begin{equation} 
\psi^{a}(\theta) = \exp\left[- \frac{1}{2} \left( 
\sum_{j}\theta^{j}H^{a}_{j b} + {\tilde W}_{\theta}\delta^{a}_{b}
\right) \right] q^{b}\ , \label{eq:exp} 
\end{equation} 
where ${\tilde W}_{\theta} = W_{\theta}-W_{0}$ and $q^{a} = 
\psi^{a}(0)$ is the prescribed state at $\theta^{j}=0$. 

The real Hilbert space ${\cal H}$, however, is not what we view here 
as the true state space of the physical model under study since 
there is an extra degree of freedom, namely, the overall 
normalisation. That is to say, the expectation of any physical 
observable is independent of the value of the normalisation. Hence 
we can gauge this extra degree of freedom away by identifying all 
the points in ${\cal H}$ along the given ray that passes through the 
origin of ${\cal H}$, corresponding to different normalisation 
factors. The resulting space obtained is the real projective 
$n$-space $RP^{n}$ which we view as the actual state space, with the 
state vectors $\psi^{a}$ in ${\cal H}$ also representing the 
homogeneous coordinates for $RP^{n}$. Note that we encounter an 
analogous situation in quantum mechanics where the overall phase 
degree of freedom associated with a given quantum state can be 
eliminated \cite{kibble} to recover the 
quantum phase space $CP^{n}$. \par 

Now, given the exponential states (\ref{eq:exp}), we may wish to ask 
if there is any symmetry associated with the flow induced by changing 
the canonical parameters $\theta^{j}$ in $RP^{n}$. This line of 
enquiry has been investigated recently \cite{dblh3}, and it was shown 
that the induced flow gives rise to a projective automorphism on the 
state manifold---a general transformation of the kind that projects 
geodesics onto geodesics. Furthermore, it has also been shown that 
the induced flow is a Hamiltonian gradient flow with respect to the 
natural `spherical' metric on the state space. In other words, the 
field generated by the tangent vectors of the homogeneous coordinates 
$\zeta_{j}^{a} = \partial_{j}\psi^{a}$ is given by the gradient 
\begin{equation} 
\zeta_{j}^{a}\ =\ -\frac{1}{4}\nabla^{a}H_{j}\ , \label{grad}
\end{equation} 
where $H_j$ is a globally defined function on $RP^{n}$, given by the 
expectation value $H^{a}_{j b}\psi^{b}\psi_{a}$ of the Hamiltonian 
operator $H_{jab}$ in the state $\psi^{a}$, and $\nabla^{a} = 
g^{ab}\nabla_{b}$ denotes the gradient operator with respect to the
homogeneous coordinates. This implies that if we consider the 
coordinate transformation induced by changing each of the coupling 
constants $\theta^{j}$ according to the differential equation 
(\ref{eq:thermo}), then in local coordinates the vector field 
$\zeta^{a} = \partial_{j}\psi^{a}$ for each given $j$, in local 
coordinates, satisfies the general equation for projective 
transformations: 
\begin{equation} 
\ze^{c}_{;a;b} + R^{c}_{bad}\ze^{d}\ =\ \delta^{c}_{(a}\phi_{b)}\ , 
        \label{proj}
\end{equation} 
where $\phi_{a}$ is a covector given by $\phi_{a} = 
(\ze^{b}_{;b})_{;a}$, $R^{c}_{bad}$ is the curvature tensor, and the 
expression $X_{;a}$ denotes the covariant derivative of $X$ with 
respect to the natural spherical metric. \par 

Having in mind the fairly general results noted above on the symmetry 
associated with the parameter development of the equilibrium states 
$\psi^{a}$, we would now like to specialise further and consider the 
flow induced by real-space renormalisation transformations. In 
particular, we now view the coupling constants $\theta^{j}(t)$ 
themselves as dependent on a single scale parameter $t$. As a 
consequence, we may view the gradient flow in (\ref{grad})
as a one--parameter family of flows, 
corresponding to each given scale. In this case, it 
is not difficult to show that the induced flow due to a change of
scale is not quite a Hamiltonian gradient flow in an ordinary sense, and 
we require some appropriate modifications. \par 

To develop this further, we simply apply the chain rule in the 
equation (\ref{eq:thermo}) for the parameter development of the 
state vector. Hence, if we write ${\dot \psi}^{a}$ for 
$\partial\psi^{a}/\partial t$, we have ${\dot \psi}^{a} = 
-\frac{1}{2}{\dot \theta}^{j}{\tilde H}^{a}_{j b}\psi^{b}$. 
Furthermore, by noticing that $\nabla^{a}H_{j} = 
2{\tilde H}^{a}_{j b}\psi^{b}$, we can write the defining equation for
the integral curve
induced by renormalisation transformations in the form 
\be
 \frac{\ptl \ps^a}{\ptl t} & = & -\frac{1}{4}\del^a\Ph  
       +\frac{1}{4}H_i\del^a\beta^i
\ee
where the potential function $\Phi$ is given by the contraction 
$\Phi = \beta^{i}H_{i}$ of the Hamiltonian function with the RG 
$\beta$-functions, given by 
$\beta^{i} = \partial\theta^{i}/\partial t$. One observes that
this integral curve does not in general correspond to a gradient flow
of the form (\ref{grad}). However, we note that the potential function
$\Ph$ is scale dependent, i.e. $\Ph=\Ph(t)$, and is a global function
on $RP^n\times${\bf R}$_+$. If we define the restriction
of this function to the state space at a constant but arbitrary scale
$t=t_0$ as $\Ph_{t_0}$, then the generating vector field for the flow,
restricted to the given constant scale, takes the form
\begin{equation} 
 \left.\frac{\ptl \ps^{a}}{\ptl t}\right|_{t_0}\ =\ 
       -\frac{1}{4} \del^{a}\Phi_{t_0}\ . 
\end{equation} 
From this expression, 
we can make the following observation, namely, the flow induced by 
the renormalisation transformations on the state space manifold 
$RP^{n}$ induces, at {\it each given scale}, a projective vector 
field \cite{dblh3}. However, the integral 
curves associated with such a flow cannot 
be obtained from any global projective vector field, since the flow 
changes with respect to the given scale characterised by the variable 
$t$. \par 

The situation just described is quite analogous
to the mechanics of a time dependent 
Hamiltonian \cite{am}. In this case, the Hamiltonian 
`gradient' flow gives rise to, at each instant of time, a Killing 
vector field. However, the integral curve along any given time 
evolution does not correspond to a global isometry of the manifold, 
since the Killing field changes in time. \par 
 
The general property of the induced flow in the state space 
that we have observed is worth bearing in mind when we specialise to 
study the flow restricted to the parameter space itself. In 
particular, since the parameter space manifold is a subspace of the 
state-space, it is natural to consider how the above structure might 
naturally project down onto the parameter space. In order to address 
this question we now study an explicit example, namely, the 
one-dimensional Ising model in an applied magnetic field. In this 
case the space of coupling constants is two-dimensional, a 
submanifold of the possibly infinite dimensional state space, 
endowed with the natural Fisher-Rao metric. \par 

\section{Parameter Space Geometry for the Ising Chain}

Consider the dimensionless Hamiltonian for the one-dimensional Ising 
model, given  by 
\begin{equation} 
H\ =\ - K\sum_{i=1}^{N}\sigma_{i}\sigma_{i+1} - 
h\sum_{i=1}^N\sigma_{i}\ . 
\end{equation} 
The probability distribution for the energy $H$ is given, as 
usual, by the Gibbs measure $p(H) = \exp(-H)/Z$, where $Z = 
{\rm Tr}[\exp(-H)]$ is the partition function. Here, the symbol 
${\rm Tr}$ denotes summation over the energy levels 
corresponding to all possible configurations. \par

The totality of the space of such probability distributions 
$\{p(H)\}$ can be mapped into a real Hilbert space ${\cal H}$, as 
discussed earlier, by taking the square root, whereupon we obtain 
a state vector for the system $\ps^a(H)$, which satisfies the 
normalisation condition $g_{ab}\ps^a\ps^b= 1$, determined by the 
Hilbert space metric. This implies that $\psi^a$ is an element on 
the unit sphere ${\cal S}$ in ${\cal H}$. The relevant parameter 
space submanifold ${\cal M}$ on the sphere is determined by 
$\psi^{a}(\theta)$ where $\{\th^i\} = (K,h)$ are local 
parameters. \par 

A Riemannian metric on ${\cal M}$, induced by the spherical geometry 
of ${\cal S}$ \cite{dblh2}, is given by the Fisher-Rao metric which, 
in local coordinates, is expressed as
\begin{equation} 
G_{ij}\ =\ 4g_{ab}\ptl_{i}\psi^a\ptl_{j}\psi^b , 
\end{equation} 
where $\partial_{i} = \partial/\partial\theta^{i}$ denotes 
differentiation with respect to the coordinates $(\theta^{1}, 
\theta^{2}) = (K,h)$. In particular, when the distribution is 
of Gibbs type, the Fisher-Rao metric reduces to the 
form $G_{ij} = -\ptl_{i}\ptl_{j} \ln Z, \label{met}$ 
which does not explicitly involve the trace operation. \par 

In the case of an $N$-spin Ising chain, the components of the 
Fisher-Rao metric can be calculated explicitly and are given by 
\begin{eqnarray}  
\left\{ \begin{array}{l} 
 G_{11}^{(N)}\ =\ G_{11} + \sum_{n=1}^{\infty} (-1)^n
   \left(\frac{c-\eta}{c+\eta}\right)^{nN} \left[
  \frac{8c(1+e^{4K}-e^{8K}+e^{8K}c_2 
   -e^{4K}nNc\sqrt{2c_2-2+4e^{-4K}})}
       {(e^{4K}-1)^2(1+e^{4K}s^2)\eta}\right]\ ,\\ 
 G_{12}^{(N)}\ =\ G_{12} + \sum_{n=1}^{\infty} (-1)^n
      \left(\frac{c-\eta}{c+\eta}\right)^{nN} 
  \left[\frac{8s(-1+e^{4K}c^2-e^{4K}s^2+2e^{4K}nNc\eta)}
     {(e^{4K}-1)^2(2-e^{4K}+e^{4K}c_2)\eta}\right]\ ,\\
 G_{22}^{(N)}\ =\ G_{22} - \sum_{n=1}^{\infty} (-1)^n
  \left(\frac{c-\eta}{c+\eta}\right)^{nN} 
  \left[\frac{2(-c+2e^{4K}nNs^2\eta)} {(1+e^{4K}s^2)\eta}\right]\ , 
\end{array} \right. \label{eq:G^N} 
\end{eqnarray}  
where $\eta = \sqrt{\sinh^2 h+e^{-4K}}$, and we use the notation 
$c\equiv \cosh h$, $s\equiv \sinh h$, and $c_2\equiv \cosh 2h$. 
The components $G_{ij}$ correspond to the result obtained when we 
take the thermodynamic limit $N\rightarrow\infty$ directly for the
ratio $G_{ij} = \lim_{N\rightarrow\infty} \partial_{i}\partial_{j} 
\ln Z/N$. The resulting expression for the metric, obtained by 
taking this limit, is given by \cite{janyszek89} 
\begin{eqnarray} 
\left\{ \begin{array}{l} 
 G_{11}\ =\ (8s^2(c+\eta)+4e^{-4K}c)
e^{-4K}\eta^{-3}(c+\eta)^{-2}\ ,\\
 G_{12}\ =\ 2s e^{-4K}\eta^{-3}\ , \\
 G_{22}\ =\ c e^{-4K}\eta^{-3}\ . \end{array} \right. 
\end{eqnarray}   
As we have noted above, consideration of the Riemannian structure in 
the thermodynamic limit requires care, as the standard definition of 
the Fisher-Rao metric then requires a suitable integration measure 
over an infinite dimensional space. That is to say, the Gibbs measure 
may not be well defined as such when we have an infinite number of 
degrees of freedom. Although it appears that one may take an appropriate limit 
in the present case, some caution is required in interpreting the 
geometry. Indeed, as we shall discuss shortly, the geometry for 
finite $N$ differs markedly from that observed in the thermodynamic 
limit. \par 

We first note that the components of the Christoffel symbols arising 
from the metric in this case are given by the simple expression 
$\Gamma^{i}_{jk} = \frac{1}{2}G^{il}\partial_{l}G_{jk}$, and hence 
the Riemann tensor reduces to the form 
\begin{equation} 
R^{i}_{jkl}\ =\ \Gamma^{i}_{ml}\Gamma^{m}_{jk} - 
\Gamma^{i}_{mk}\Gamma^{m}_{jl}\ . 
\end{equation} 
In particular, in two dimensions it suffices to consider just the 
scalar curvature, given by $R = 2R_{1212}/{\rm det}(G_{ij})$. Hence, 
by evaluating the nonzero component of the Riemann tensor in the 
thermodynamic limit, we find 
$R_{1212} = 2[e^{8K}\gamma^{5}(\cosh h+\eta)]^{-1}$, from which it 
follows \cite{janyszek89} that 
\begin{equation} 
R\ =\ 1 + \eta^{-1} \cosh h \ , \label{Rs}
\end{equation} 
which is always positive. \par 

However, if we consider the parameter space geometry for finite $N$ 
using the expressions (\ref{eq:G^N}), then we observe that when the 
size of the system $N$ is below certain critical values $N_{c}(K,h)$, 
the scalar curvature is negative. This is illustrated in Fig.~1 where 
the scalar curvature is plotted against the system size $N$ for the parameter 
values $K=h=1$. As $N$ increases the curvature tends 
asymptotically to the thermodynamic value $R$, while for $N < N_{c}$ 
($N_{c}=7$ in this case) the curvature is negative, indicating a 
radically different parameter space geometry.
\begin{figure}
\label{rn}
 \centerline{%
   \psfig{file=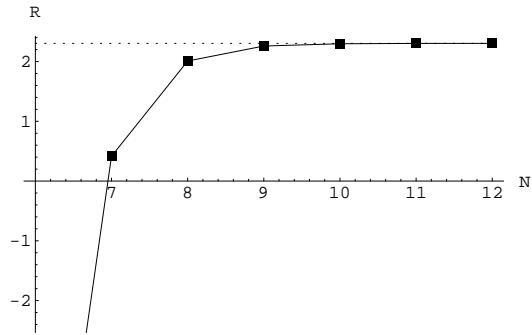,width=7cm,angle=0}%
   }
 \caption{A plot of the Ricci scalar $R(N)$ at the point $K=h=1$. The 
 value in the thermodynamic limit $R \approx 2.30$ is approached as
 $N$ becomes large. } 
\end{figure}
Nevertheless, since we are only interested in the limit of large
$N$, and as this limit appears to be well defined, we shall generally 
work henceforth with the Riemannian structure obtained after taking 
thermodynamic limit. \par 

We conclude this section with a brief discussion of the physical
significance of the Ricci curvature. This is a natural question to 
address since the curvature can be expressed \cite{stat} in terms of 
a combination of higher order central moments of the Hamiltonian $H$,
while geometrically it is an invariant quantity with respect to 
reparametrisation. The expression based on the central moments of 
$H$ allows a study of the scaling behaviour of $R$ in the vicinity of 
the critical points for those models exhibiting phase transitions, 
from which Ruppeiner \cite{stat} conjectured that the curvature is a 
measure of the correlation volume of the system (see also 
\cite{janyszek89}). \par 

In the present case we can compare $R$ directly with the correlation 
length $\xi$ of the Ising chain. If we write $\cot(2\phi) = 
e^{2K}\sinh(h)$, then the correlation function $C(r) = 
\langle\sigma_{i}\sigma_{i+r}\rangle - 
\langle\sigma_{i}\rangle\langle\sigma_{i+r}\rangle$ can be expressed, 
after taking the limit $N\rightarrow\infty$, in the simple form 
\begin{equation} 
C(r)\ =\ \sin^{2}(2\phi)\left(\frac{\lambda_{+}}{\lambda_{-}} 
\right)^{r}\ , 
\end{equation} 
where $\lambda_{\pm} = e^{K}(c \pm \et)$ are the eigenvalues 
of the transfer matrix $V$ discussed below in Section 4. Therefore, 
for the correlation length $\xi$ we find that 
\begin{equation} 
\xi\ =\ \frac{1}{\ln(\lambda_{+}/\lambda_{-})}\ =\ 
         -\ln\left(1-\frac{2}{R}\right)\ ,  \label{xi}
\end{equation} 
where we have used (\ref{Rs}) to determine the exact relation between 
$\xi$ and the Ricci scalar $R$. Using standard techniques for 
inverting a power series, this relation may be expressed in the form, 
\begin{equation} 
\xi\ =\ \sum_{k=0}^{\infty} c_k \left(\frac{2}{R}\right)^k\ =\ 
\frac{R}{2}\left(1-\frac{1}{R}-\frac{1}{3R^2} - 
O\left(\frac{1}{R^3} \right) \right)\ ,
\end{equation} 
where the coefficients $c_k$ are given by $c_{k} = - \sum_{p=1}^{k} 
\frac{1}{p+1}c_{k-p}$ with the initial value $c_0=R/2$. From this 
relation we observe that near the critical point $\xi\sim R/2$ and 
thus $R/2$ provides a good quantitative measure of the correlation 
length. Such a connection breaks down, however, in any regime
where $R\sim 1$.\par 

Having discussed the Riemannian geometry associated with the 
parameter space of the 1D Ising model, we shall, in the next 
section, determine the exact diffeomorphism of this manifold 
induced by an infinitesimal  rescaling of the lattice spacing. 
Subsequently, we shall then consider the symmetry properties of 
this mapping. \par 

\section{Calculation of the RG $\beta$--Functions}

The standard approach to real-space renormalisation is implemented by 
means of a discrete scaling transformation, known as block-spin 
decimation. In the case of the 1D Ising chain this transformation can 
be performed exactly. However, such transformations are not 
advantageous in seeking differentiable structures on the parameter 
space, such as we are investigating here. Therefore, in order to extract a 
generating vector field induced by these transformations, an 
appropriate form of analytic continuation is required. In this 
section we present a relatively simple technique based on the 
transfer matrix for achieving this aim.\par 

We consider first a discrete decimation of the $N$-spin system 
via the blocking of lattice sites by an integer factor $l$ at each stage of 
the decimation process. Thus, if we denote by 
$a$ the lattice spacing between the spins, then at each iteration 
the lattice size increases by a factor of $l$, that is, 
$a \rightarrow la$. In addition, we assume a periodical boundary 
condition on the lattice. The renormalised parameters $K'$ and $h'$, 
after rescaling, are determined implicitly by the relation 
\begin{equation} 
Z_{N/l}(K',h')\ =\ f Z_N(K,h), \label{rgreln}
\end{equation} 
where $f$, a function of $K$ and $h$, is an overall scaling factor. 
Representing the partition function by $Z_N = {\rm Tr}(V^N)$ in terms 
of the transfer matrix $V$, given by 
\begin{equation} 
V\ =\ \left(\begin{array}{cc}
       e^{K+h}   &   e^{-K} \\
       e^{-K}    &   e^{K-h} 
               \end{array}\right)\ ,
\end{equation} 
we can reexpress the RG relation (\ref{rgreln}) in terms of individual
configurations as 
\begin{equation} 
V(K',h')\ =\ f^{l/N} V^{l}(K,h)\ .
\end{equation} 
In particular, we see that the following relations 
\begin{eqnarray} 
 e^{2h'}\ =\ \frac{(V^{l})_{11}}{(V^{l})_{22}}\ , \ \ \ \ 
 e^{4K'}\ =\ \frac{(V^{l})_{11}(V^{l})_{22}}{(V^{l})_{12}^2} 
      \label{eq:rec} 
\end{eqnarray} 
hold. Therefore, in order to extract the recursion relations we 
require the transfer matrix raised to an arbitrary power $l$. This 
is achieved by the use of a similarity transformation $S$ to 
diagonalise the transfer matrix $V$. Since the eigenvalues of the 
matrix $V$ are given by 
\begin{equation} 
\lambda_{\pm}\ =\ e^K(\cosh h \pm \eta)\ ,
\end{equation} 
if we denote the diagonalisation of $V$ by $\Lambda$, i.e. 
$V = S\Lambda S^{-1}$, we find 
\begin{equation} 
 V^{l}\ =\ S\left(\begin{array}{cc}
       \lambda_{+}^{l}   &   0 \\
       0    &   \lambda_{-}^{l} 
               \end{array}\right)S^{-1}.
\end{equation} 
Recalling the definition $\cot(2\phi) = e^{2K}\sinh h$, this 
similarity transformation can be expressed in a simple form: 
\begin{equation} 
S\ =\ \left(\begin{array}{cc}
       1   &   -\tan\phi \\ 
       \tan\phi    &   1
       \end{array}\right)\ .
\end{equation} 
Thus, from the expressions in (\ref{eq:rec}), the recursion 
relations for the rescaled coupling constants can be obtained as  
\begin{equation}  
 e^{2h'}\ =\ \frac{(c+\eta)^{l}+(c-\eta)^{l}\tan^2\ph}
     {(c-\eta)^{l}+(c+\eta)^{l}\tan^2\ph} \\ \label{rec1}\ , 
\end{equation} 
\begin{equation} 
 e^{4K'}\ =\ 1+\frac{4e^{4K}\eta^2(1-e^{-4K})^{l}}
        {[(c+\eta)^{l}-(c-\eta)^{l}]^2}\ . \label{rec2}
\end{equation} 
Here, for clarity, we recall the notation $s\equiv \sinh h$ and 
$c \equiv \cosh h$. One readily verifies that these expressions 
reproduce the standard recursion relations for 2-spin blocking if we 
set $l=2$, and also the recursion relation for arbitrary $l$, when 
$h=0$, as discussed in \cite{kadanoff77}.\par 

In the present context we assume that we can analytically continue 
these results to arbitrary $l \in {\bf R}_{+}$. It is convenient 
therefore to proceed by setting $l =1+\epsilon$, where $\epsilon 
\ll 1$. Then in expressions (\ref{rec1}) and (\ref{rec2}) the 
parameters now have a full lattice spacing dependence of the form 
$\theta = \theta(a/a_0)$ and $\theta' = \theta'(a/a_0+\epsilon a/a_0)$ where 
$\theta = (K,h)$ and $a_0$ is the short distance cutoff scale. 
We then perform a Taylor series expansion of the 
left hand side of (\ref{rec1}) and (\ref{rec2}) to
$O(\epsilon)$. The right hand side can be expanded to the same order 
using the standard relation $x^{\epsilon}\approx 1+\epsilon\ln x + 
O(\epsilon^2)$. Equating terms of $O(\epsilon)$ we obtain the 
$\beta$--functions for the parameters $K$ and $h$. In particular, 
it is convenient to write the lattice spacing as $a/a_0=e^t$, 
so that the expressions $\beta^i\equiv a\partial 
\theta^{i}/\partial a$ take the form, 
\begin{equation} 
 \beta^K\ \equiv\ \frac{\partial K}{\partial t}\ =\ 
\frac{e^{-2K}\sinh 2K\cosh h}{2\eta} \ln 
\frac{\cosh h-\eta}{\cosh h+\eta} \label{betk} 
\end{equation} 
\begin{equation} 
 \beta^h\ \equiv\ \frac{\partial h}{\partial t}\ =\ 
\frac{e^{-2K}\sinh 2K\sinh h}{\eta}\ln 
\frac{\cosh h+\eta}{\cosh h-\eta}\ . \label{beth} 
\end{equation} 
Note that $\beta^K$ and $\beta^h$ have a simple proportionality 
relation $\beta^h= -2\tanh h \beta^K$. From these expressions one may 
readily verify various limiting cases. In particular, for arbitrary 
$h$ there is a line of trivial fixed points at $K=0$. In the limit 
$h\rightarrow 0$, the first component $\beta^K$ reduces to the 
familiar form $\beta^{K}_{h=0} = \frac{1}{2}\sinh 2K \ln(\tanh K)$, 
while if we work to $O(h)$ we also find that $\beta^{h} = -2h\beta^{K}_{h=0} 
+ O(h^3)$, as discussed in \cite{bl}.
\begin{figure}
\label{vect}
 \centerline{%
   \psfig{file=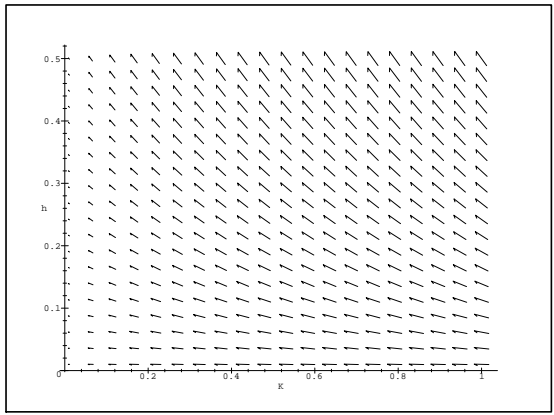,width=8cm,angle=270}%
   }
 \caption{A field plot of the differential RG generating 
vector field $\xi^i$ in the $(K,h)$ plane, associated with the 
one-dimensional Ising model.} 
\end{figure}

With these generating functions at hand we can view the 
infinitesimal RG transformations as inducing the following 
diffeomorphism on the parameter space manifold, 
\begin{equation} 
\theta^{i}\ \longrightarrow\ \theta^{i} + \epsilon \xi^{i}\ , 
\end{equation} 
where $\epsilon$ is an infinitesimal parameter and the generating 
vector field $vec{\xi}$ is given by the RG $\beta$-functions, 
\begin{equation} 
\vec{\xi}\ =\ \beta^i\frac{\ptl}{\ptl\th^i}.
\end{equation} 
The behaviour of this vector field is displayed in Fig.~2, indicating 
the flow of the RG trajectories away from the critical point at 
$K=\infty$ towards the line of stable high temperature fixed points 
at $K=0$. \par 

In the next section we shall study the geometrical properties of this 
vector field in various limiting cases. Before proceeding to do so, 
we note that certain physical features of the model contained 
in the vector field $\xi^i$ are only made explicit when the underlying 
Riemannian structure is taken into account. To this end, we consider 
the conventional definition of the fixed points, as identified with 
the zeros of the $\beta$-functions. This definition is not entirely 
satisfactory, since there are cases where the $\beta$-function does 
not vanish at the critical point, such as the example considered 
here. While the $\beta$-functions vanish at one of the fixed 
points, $K=0$, they tend to constant values for any given value of 
$h$ when $K\rightarrow\infty$. \par 

Clearly, if the underlying geometry were Euclidean, then this implies 
that ($K=\infty,h=0$) cannot be regarded as a proper fixed point. 
However, the point here is that the parameter space is a curved 
manifold endowed with the Fisher-Rao metric. Therefore, any fixed 
point of a given vector field generating a diffeomorphism on the 
manifold, in general, can be (cf. \cite{brody87}) identified 
with the zeros of the `velocity function' defined by 
\begin{equation} 
v\ =\ \sqrt{G_{ij}\beta^i\beta^j}\ .  
\end{equation} 
The behaviour of the velocity function in the case of the 1D Ising 
model is illustrated in Fig.~3. It is clear that the proper velocity 
of the flow vanishes at both fixed points. An alternative 
interpretation of this point is available in two dimensions where 
the existence of the Zamolodchikov $C$--function allows the $\beta$ 
function to be constructed as $\beta^i=G^{ij}\ptl_j C$ near the fixed 
points \cite{zamol}. When the renormalization group 
generates a gradient flow of this form \cite{grad}, it is then clear that
the fundamental covector $\beta_j=\ptl_j C$ also encodes the correct fixed 
point structure of the flow.
\begin{figure}
\label{vel}
 \centerline{%
   \psfig{file=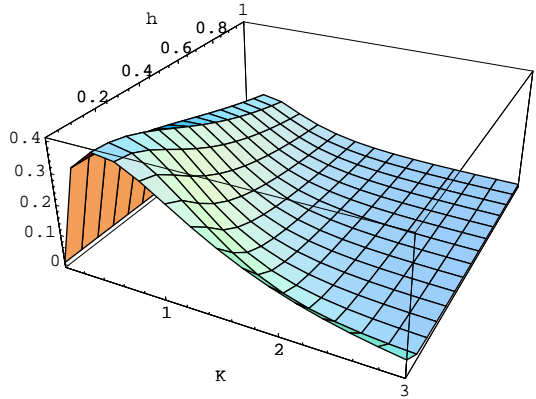,width=7cm,angle=0}%
   }
 \caption{A plot of the velocity function $v(K,h)$ over the 
parameter space of the one dimensional Ising model. The velocity of the flow
vanishes for an arbitrary value of $h$ at the high temperature fixed points at 
$K=0$, and also at the critical point
given by $K=\infty$ and $h=0$. } 
\end{figure}

\section{The Symmetry of RG Diffeomorphisms} 

As we have discussed in Section 2, the RG generating vector field 
in the projective state space may be interpreted 
as generating a projective automorphism of the 
{\it fixed scale} state-space manifold. On the other hand, if 
we consider the global flow induced by the RG, then there seems to 
be no generic symmetry associated with the flow that can be 
expressed in terms of conventional diffeomorphisms in 
Riemannian geometry. This fact then leads us to conjecture that 
such a symmetry should also be absent for RG flow viewed in its 
normal setting of the parameter space submanifold. 
In this case the restriction of the natural state-space metric to 
the parameter space can be shown \cite{dblh3} to coincide
with the Fisher-Rao metric which, as we have discussed in 
Section 3, can be explicitly evaluated in local coordinates 
which are parameters of the Hamiltonian. The components of the 
generating vector field for the full scale-dependent flow then reduce to the 
$\beta$--functions for these parameters. Thus we are led to 
enquire whether the vector field $\xi^i$ is a projective vector 
field on the parameter space manifold. Indeed, this is also 
a natural question on general grounds from the standard theory 
of Riemannian geometry, since projective automorphisms are the 
most general diffeomorphisms preserving geodesics on a manifold, 
and include affine and Killing transformations as special cases. \par 

In fact, as we shall observe below, by explicit analysis of a counter 
example---the one dimensional Ising model---the flow induced by the 
RG in general does not possess any {\it standard} global symmetry. In 
particular, the exact generating vector field for RG flow in the 
1D Ising model, $\xi^i$, is not a projective vector field. This is of 
course consistent with the observations in state space where there is 
an explicit correction term. \par

This statement is verified in the case of the 1D Ising model 
in a straightforward manner since for $\xi^i$, evaluated in Section 4, 
to be a projective vector field it must satisfy the relation 
(\ref{proj}),
\be
 \xi^k_{;j;i} + R^k_{jil}\xi^l & = & \de_{(j}^k\ph_{i)},
\ee
for some covariant vector $\ph_i$. In two dimensions, the 
components of this equation are explicitly given by 
\begin{eqnarray}  
\left\{ \begin{array}{ll} 
 \xi^1_{;2;2}+G^{11}R_{1221}\xi^1 = 0 
   & \xi^2_{;1;1}+G^{22}R_{2112}\xi^2 = 0, \\
 \xi^1_{;1;1}+G^{12}R_{2112}\xi^2 = 2\ph_1 
   & \xi^2_{;2;2}+G^{21}R_{1221}\xi^1 = 2\ph_2, \\
 \xi^1_{;1;2}+G^{12}R_{2121}\xi^1 = \ph_2 
   & \xi^2_{;2;1}+G^{21}R_{1212}\xi^2 = \ph_1, \\
 \xi^1_{;2;1}+G^{11}R_{1212}\xi^2 = \ph_2
   & \xi^2_{;1;2}+G^{22}R_{2121}\xi^1 = \ph_1,
\end{array} \right. 
\end{eqnarray} 
and by direct substitution of the expressions for the metric, 
curvature, and the vector field $\xi^{j}$, we find that 
the equation is not satisfied globally on the parameter space for 
any choice of $\ph_i$ for the geometry at finite $N$ and also in 
the thermodynamic limit. Thus $\xi^{j}$ is not a projective vector 
field on the parameter space manifold. Since we have
knowledge of the exact geometric quantities, and the RG $\beta$--functions, 
in this case the 1D Ising model in an external field serves as a 
counter-example to any conjecture of the general validity of such 
a symmetry, as stated above. \par 

It is interesting to note that as we are dealing in this case with 
a two-dimensional parameter space, this allows additional control 
through the fact that one may coordinatise the surface with a 
complex Riemann coordinate and consider general conformal 
transformations. However, as we wish to consider this model purely 
as a tractable example of higher dimensional (generically infinite 
dimensional) cases we shall avoid for the moment any aspects which 
appear particular to two dimensions. \par 

Nevertheless, we can consider various limiting cases by restricting 
attention to local regions of the parameter space. The most relevant 
region is of course the vicinity of the 
critical point. Therefore, it would be convenient to linearise the 
vector field around $K=\infty$ and $h=0$. However, as we notice from 
the relations (\ref{betk}) and (\ref{beth}), the generic dependence 
on the temperature parameter $T=1/K$ is of the form $\exp(\pm 2/T)$,
implying that $T=0$ is an essential singularity. Thus, following
standard practice for the study of critical exponents, it is 
convenient to introduce a new variable $\tau = \exp(-2K)$, and 
consider a linearisation of the RG transformation in the variables 
$(\tau,h)$ about $(0,0)$. Assuming that $\tau$ and $h$ are of 
$O(\epsilon)$, and expanding equations (\ref{betk}) and (\ref{beth}) 
up to $O(\epsilon^4)$, we obtain 
\begin{equation} 
\beta^K\ =\ -\frac{1}{2}+\left(\frac{\tau^2}{3}-\frac{h^2}{6}\right) 
+ \left(\frac{2\tau^2h^2}{15}+\frac{\tau^4}{15}+\frac{h^4}{90}\right)
+ O(\epsilon^5)\ , 
\end{equation} 
\begin{equation} 
\beta^h\ =\ h - \left(\frac{2\tau^2h}{3}\right) + O(\epsilon^5)\ . 
\end{equation} 
Note that various geometric quantities, e.g., the curvature, 
remain singular at the critical point in these variables. However, 
in terms of $\ta$ and $h$ we can construct a Laurent expansion 
and consider the structure of the lowest order terms. \par 

Rather than considering general projective automorphisms in the
neighbourhood of the critical point, it is now convenient to analyse 
particular transformations which have had some attention in the 
literature. In particular, we consider the degree to which RG flow 
violates the geodesic equation, and whether there are any special 
flows which are exact geodesics of the metric. We also investigate 
the extent to which $\xi^i$ may be regarded as a conformal Killing 
vector near the critical point. Conformal Killing transformations, 
preserving the angles between vectors, are not a subset of 
projective automorphisms but one readily verifies that, for the 1D 
Ising model, the differential RG does not generate such a symmetry 
globally. \par

We first analyse the relationship between RG diffeomorphisms and 
geodesic flow on the parameter manifold. In fact, using the standard 
discrete decimation procedure, Dolan \cite{diosi} suggested that the 
induced flow might be a geodesic along the line $h=0$. Although an 
explicit solution for the geodesic equations has not been obtained, 
we can nevertheless determine in which regimes the vector field 
$\xi^{i}$ induced by the infinitesimal scale change satisfies the 
geodesic equation $\xi^{j}\nabla_{j}\xi^{i} - A\xi^{i} = 0$, where 
$\nabla_j$ denotes the covariant derivative compatible with the 
underlying Fisher-Rao metric $G_{ij}$. The function $A$ in this 
expression vanishes only if the parameter $t = \ln a/a_0$ is an affine 
parameter. 

Numerical analysis indicates that along the lines $h=0$ and $K=0$ 
the flow is an exact geodesic flow, and that the deviation becomes 
linear in $h$, as $K$ increases. Hence our results confirm the 
expectation in \cite{diosi}. In order to see this explicitly, we 
expand the geodesic equations in $\tau$ and $h$ to obtain 
\begin{equation} 
 \xi^j\nabla_j\xi^1 - A\xi^1\ =\ 0 + 
\frac{2A-1}{4}+\frac{h}{2} +O(\epsilon^2)\ , 
\end{equation} 
\begin{equation} 
 \xi^j\nabla_j\xi^2 - A\xi^2\ =\ 0 + 
(1-A)h + O(\epsilon^2)\ .
\end{equation} 
The common factor of $A$ is given by $A=1/2+O(\epsilon)$, and the 
deviation from geodesic flow is clearly observed to be linear in $h$ 
with a positive coefficient of $1/2$ for each component.\par

We turn now to an investigation of whether $\xi^i$ corresponds to 
a conformal Killing vector field in particular regions of the 
parameter space. Justification for the conjecture that $\xi^i$ may 
have such a structure at least locally on the manifold in the 
neighbourhood of the critical point is provided by the analysis of 
Di\'osi et al. \cite{diosi} who have shown that one recovers the 
standard critical exponents of a system such as the 1D Ising model 
under the assumption that $\xi^{i}$ linearised around the critical 
point is a conformal Killing vector field. In other words, the 
components of the metric tensor $G_{ij}$ should satisfy 
${\cal L}_{\xi}G-dG =0$ where ${\cal L}_{\xi}$ denotes the Lie 
derivative with respect to the vector field $\vec{\xi}$, and $d$ is 
the spacetime dimension. \par 

As we have noted earlier, explicit evaluation of this equation in 
the present example indicates that the symmetry does not hold 
globally on the parameter space. However, $\xi^i$ is indeed a 
conformal Killing field to a good approximation near the critical 
point $K=\infty$, $h=0$. Indeed significant deviations are only seen 
as one moves close to the high temperature fixed points where $K=0$. 
\par

The approximate symmetry near the critical point may be shown more
quantitatively by expanding the conformal Killing equation in a 
Laurent expansion in the variables $\ta$ and $h$. For example, the 
$(1,1)$ and $(2,2)$ components take the form 
\begin{equation} 
 ({\cal L}_{\xi}G)_{11}-dG_{11}\ =\ 0 +  \left(\frac{8(1-d)}{h} - 
8(2-d) \frac{8(5-d)h}{3}+O(h^2)\right)\tau^{2} + O(\epsilon^4)\ , 
\end{equation} 
\begin{equation} 
 ({\cal L}_{\xi}G)_{22}-dG_{22}\ =\ 0 + \left(\frac{(1-d)}{h^3} - 
\frac{2}{3h} \frac{(11+3d)h}{45}+O(h^2)\right)\tau^2 + 
O(\epsilon^4)\ , 
\end{equation} 
and thus $\xi^i$ in fact corresponds to a Killing vector field
up to $O(\tau)$, and to a conformal Killing vector field with $d=1$ 
up to $O(\ta^2)$ for the $(1,1)$ component and up to $O(\ta^2/h)$ for 
the $(2,2)$ component. We note that this expansion is not a strict 
linearisation of the RG equations about the critical point, due to 
their singular structure. Nevertheless these results verify, in this 
example, the claim of Di\'osi et al. \cite{diosi} explicitly.\par

\section{Discussion}
We have presented an analysis of the Riemannian parameter space 
geometry of the 1D Ising model, and the characteristics of 
renormalisation group trajectories on this manifold. While the 
generator of this transformation is, up to higher order corrections, 
a conformal Killing vector field near the critical point, there is 
apparently no such interpretation globally on the manifold. This is 
consistent with the general structure observed for RG flow in the 
state space of the system. \par 

It should be pointed out, however, that in other models of interest 
such symmetries may exist due to particular properties of the theory 
itself or the individual parameters. The question of the existence of 
other independent parameter space symmetries is also of importance 
as these may provide additional, generically nonperturbative, 
constraints on the $\beta$--functions of the theory. Discrete 
parameter space symmetries are of this kind, while in the present 
case we observe that there is a class of diffeomorphisms of the 
parameter space manifold which preserve the RG flow in the sense 
that, for the generator $X$ of such a transformation, 
${\cal L}_{X}\et=0$ where $\et=\et_id\th^i$ is the 1-form dual to the
vector field $\xi^i$. One may verify that $\et$ has the form 
$\et=(\et_1,0)$ and thus any diffeomorphism generated by a vector 
field of the form $X=(0,x_2)$ will preserve the RG flow in the sense
described above. Such a structure is clearly quite model dependent. 
Nevertheless, the presence of a commuting flow may well have some 
more general validity. \par

\subsection*{Acknowledgements}

The authors would like to thank L.P. Hughston for stimulating 
discussions, and the financial support of D.B. by PPARC, and A.R by 
the Commonwealth Scholarship Commission and the British Council, 
is also gratefully acknowledged. \par 

\bibliographystyle{prsty}

\end{document}